\documentclass[aps, pra, reprint, superscriptaddress, onecolumn, notitlepage, tightenlines, 11pt]{revtex4-2}

\usepackage{header}

\graphicspath{{graphics/}}

\begin{document}

\title{Fast Algorithms for Permanents of Permutationally-Dominant Matrices}

\author{Ryan L. Mann}
\email{mail@ryanmann.org}
\homepage{http://www.ryanmann.org}
\affiliation{Centre for Quantum Software and Information, School of Computer Science, Faculty of Engineering \& Information Technology, University of Technology Sydney, NSW 2007, Australia}

\begin{abstract}
    We establish a fast algorithm for approximating the permanent of permutationally-dominant matrices. In particular, we obtain a fully polynomial randomised approximation scheme whose runtime is quadratic in both the dimension of the matrix and the inverse of the approximation error. Our approach is based on a polymer model representation of the permanent and importance sampling of the logarithm of its partition function via the theory of heaps of pieces of Viennot.
\end{abstract}

\maketitle

\section{Introduction}
\label{section:Introduction}

The permanent of a matrix is a fundamental quantity in combinatorics, statistical physics, and computational complexity theory, yet computing it exactly is \#P-hard~\cite{valiant1979complexity}. The natural question is then how well it can be approximated in polynomial time. For matrices with non-negative entries, Jerrum, Sinclair, and Vigoda established a fully polynomial randomised approximation scheme for the permanent~\cite{jerrum2004polynomial}. For matrices with arbitrary complex entries, however, even approximating the permanent to within a multiplicative factor is \#P-hard~\cite{aaronson2013computational}. Nevertheless, Gurvits established a randomised algorithm that approximates the permanent of a general matrix $A\in\mathbb{C}^{n \times n}$ to within additive error $\epsilon\norm{A}^n$ in time $O(n^2\epsilon^{-2})$, where $\norm{A}$ denotes the operator norm of $A$~\cite{gurvits2005complexity}.

Barvinok introduced a method for approximating evaluations of a polynomial by truncating the Taylor series of its logarithm, provided the polynomial has no zeros in a suitable region of the complex plane~\cite{barvinok2016combinatorics, patel2022approximate}. This approach yields multiplicative approximations of the permanent in quasi-polynomial time for complex matrices whose entries are close to one~\cite{barvinok2016computing, barvinok2017approximating}, $\lambda$-permutationally-dominant matrices for any fixed $\lambda>1$~\cite{barvinok2019computing}, and random matrices with vanishing mean~\cite{eldar2018approximating} (see also Ref.~\cite{ji2021approximating}). Further, it yields multiplicative approximations in polynomial time for random matrices with polynomially small mean~\cite{koehler2026approximating}. Patel and Regts obtained polynomial-time algorithms from this method for a large class of graph polynomials on bounded-degree graphs~\cite{patel2017deterministic, patel2022approximate}. However, their technique does not appear to extend to permanents of dense matrices.

The polymer model formalism provides an alternative framework for developing approximation algorithms~\cite{gruber1971general, kotecky1986cluster}. The main algorithmic tool in this framework is the cluster expansion, which has been used to obtain efficient algorithms for a range of counting problems~\cite{helmuth2020algorithmic, jenssen2020algorithms, borgs2020efficient, coulson2020statistical, mann2021efficient, helmuth2023efficient, mann2025approximate}. These algorithms compute a truncation of the cluster expansion by enumerating all clusters up to logarithmic size, and as such are limited to bounded-degree settings. Faster algorithms that avoid enumeration have been obtained via Markov chains on polymer configurations, but these require the polymer weights to be non-negative~\cite{chen2021fast, galanis2022fast, blanca2024fast, mann2026fast}.

The permanent admits a natural polymer model representation in which the polymers are cycles. For dense matrices, the number of cycles of a given length containing any element grows polynomially with the dimension, so enumerating clusters of logarithmic size requires quasi-polynomial time. Further, the polymer weights are in general complex, and so Markov chains on polymer configurations do not directly apply. Both of these limitations can be avoided by sampling terms of an expansion of the logarithm of the partition function rather than enumerating them, as in algorithms that sample terms of the cluster expansion based on importance sampling~\cite{chen2025convergence, putterman2026quantum} and Markov chain methods~\cite{sanchez2026efficient}.

In this paper, we establish a fully polynomial randomised approximation scheme for the permanent of $\lambda$-permutationally-dominant $n \times n$ matrices for any fixed $\lambda>1$, with runtime $O(n^2\epsilon^{-2})$, where $\epsilon$ is the approximation error. This is the same regime in which Barvinok's approach yields a quasi-polynomial-time algorithm~\cite{barvinok2019computing}. Our approach is based on importance sampling of the logarithm of the partition function of the cycle polymer model. In particular, we use the theory of heaps of pieces of Viennot~\cite{viennot1986heaps} to express the logarithm of the partition function as a sum over pyramids, which are obtained from closed walks via loop erasure. This pyramid expansion coincides with the cluster expansion as a formal power series, but its terms correspond to closed walks, which can be sampled in proportion to the absolute values of their weights. The resulting estimator is unbiased, with constant expected runtime after $O(n^2)$ preprocessing and second moment $O(n^2)$.

This paper is structured as follows. In Section~\ref{section:Preliminaries}, we introduce the necessary preliminaries. Then, in Section~\ref{section:PolymerModelRepresentation}, we establish a polymer model representation of the permanent and express the logarithm of its partition function as a sum over closed walks for permutationally-dominant matrices. In Section~\ref{section:ApproximationAlgorithms}, we establish our approximation scheme for the permanent. Finally, we conclude in Section~\ref{section:ConclusionOutlook} with some remarks and open problems.

\section{Preliminaries}
\label{section:Preliminaries}

\subsection{Abstract Polymer Models}
\label{section:AbstractPolymerModels}

An \emph{abstract polymer model} is a triple $\mathcal{P}=(\mathcal{C},w,\sim)$, where $\mathcal{C}$ is a finite set of objects called \emph{polymers}, $w:\mathcal{C}\to\mathbb{C}$ is a function that assigns to each polymer $\gamma\in\mathcal{C}$ a complex weight $w_\gamma$, and $\sim$ is a symmetric compatibility relation such that each polymer is incompatible with itself. A set of polymers is called \emph{admissible} if the polymers in the set are all pairwise compatible. Note that the empty set is admissible. Let $\mathcal{G}$ denote the collection of all admissible sets of polymers from $\mathcal{C}$. The \emph{partition function} of $\mathcal{P}$ is defined by
\begin{equation}
    Z_\mathcal{P} \coloneqq \sum_{\Gamma\in\mathcal{G}}\prod_{\gamma\in\Gamma}w_\gamma. \notag
\end{equation}

\subsection{Pyramid Expansion}
\label{section:PyramidExpansion}

We now define the pyramid expansion~\cite{viennot1986heaps, krattenthaler2006theory}. Two ordered tuples of polymers are \emph{equivalent} if one can be obtained from the other by repeatedly interchanging adjacent compatible polymers. A \emph{heap} is an equivalence class of non-empty ordered tuples of polymers, and a \emph{pyramid} is a heap whose tuples all end with the same polymer. Let $\mathcal{G}_\mathrm{P}$ denote the set of all pyramids of polymers from $\mathcal{C}$. The \emph{pyramid expansion} is a formal power series for $\log(Z_\mathcal{P})$ in the variables $w_\gamma$, defined by
\begin{equation}
    \log(Z_\mathcal{P}) \coloneqq \sum_{H\in\mathcal{G}_\mathrm{P}}\frac{(-1)^{\abs{H}-1}}{\abs{H}}\prod_{\gamma\in H}w_\gamma. \notag
\end{equation}
This follows from the inversion lemma and the logarithmic lemma of Viennot~\cite{viennot1986heaps}. We note that the pyramid expansion and the cluster expansion~\cite{kotecky1986cluster, friedli2017statistical} are formal power series for the same quantity, and hence coincide as formal power series.

\subsection{Permutationally-Dominant Matrices}
\label{section:PermutationallyDominantMatrices}

Let $S_n$ denote the symmetric group on $[n]$. For a cycle $\gamma \in S_n$, we denote its length by $\abs{\gamma}$ and its support by $\bar{\gamma}$. The \emph{permanent} of a matrix $A=(a_{u,v})\in\mathbb{C}^{n \times n}$ is defined by
\begin{equation}
    \operatorname{per}(A) \coloneqq \sum_{\sigma \in S_n}\prod_{u\in[n]}a_{u,\sigma(u)}. \notag
\end{equation}
Note that the permanent is invariant under permutations of the rows and columns of $A$. For $\lambda>0$, the matrix $A$ is \emph{$\lambda$-diagonally-dominant} if, for all $u\in[n]$,
\begin{equation}
    \lambda\sum_{v\in[n]{\setminus}\{u\}}\abs{a_{u,v}} \leq \abs{a_{u,u}}. \notag
\end{equation}
We consider the following generalisation of diagonally-dominant matrices.

\begin{definition}[Permutationally-dominant matrix]
    Let $\lambda>0$. A matrix $A=(a_{u,v})\in\mathbb{C}^{n \times n}$ is \emph{$\lambda$-permutationally-dominant} if there exists a permutation $\pi \in S_n$ such that, for all $u\in[n]$, the entry $a_{u,\pi(u)}$ is non-zero and
    \begin{equation}
        \lambda\sum_{v\in[n]{\setminus}\{u\}}\abs{a_{u,\pi(v)}} \leq \abs{a_{u,\pi(u)}}. \notag
    \end{equation}
\end{definition}

Note that the $\lambda$-permutationally-dominant matrices with respect to the identity permutation are exactly the $\lambda$-diagonally-dominant matrices with non-zero diagonal entries. Further, every $\lambda$-permutationally-dominant matrix with respect to a permutation $\pi$ can be made $\lambda$-permutationally-dominant with respect to the identity permutation by permuting its columns according to $\pi$, which leaves its permanent unchanged.

\subsection{Approximation Schemes}
\label{section:ApproximationSchemes}

Let $\epsilon>0$. A \emph{multiplicative $\epsilon$-approximation} to a complex number $z$ is a complex number $\hat{z}$ such that $\abs{z-\hat{z}}\leq\epsilon\abs{z}$. A \emph{fully polynomial randomised approximation scheme} for a sequence of complex numbers $(z_n)_{n\in\mathbb{N}}$ is a randomised algorithm that, for any $n$ and any $0<\epsilon\leq1$, outputs a multiplicative $\epsilon$-approximation to $z_n$ with probability at least $\frac{2}{3}$, in time polynomial in $n$ and $\epsilon^{-1}$. We work in the real-RAM model of computation, in which arithmetic operations on complex numbers, evaluations of elementary functions, and samples of uniformly random numbers in $[0,1]$ each take unit time.

\section{Polymer Model Representation}
\label{section:PolymerModelRepresentation}

In this section we express the permanent of a matrix in terms of the partition function of a polymer model, and express the logarithm of the partition function as a sum over closed walks for permutationally-dominant matrices.

Let $A=(a_{u,v})\in\mathbb{C}^{n \times n}$ be a matrix and let $\pi \in S_n$ be a permutation such that the entries $a_{u,\pi(u)}$ are non-zero for all $u\in[n]$. We define the \emph{cycle polymer model} of $A$ with respect to $\pi$ to be the abstract polymer model $\mathcal{P}=(\mathcal{C},w,\sim)$, where $\mathcal{C}$ is the set of all cycles in $S_n$ of length at least $2$, two polymers $\gamma$ and $\gamma'$ are compatible if and only if $\bar{\gamma}\cap\bar{\gamma}'=\varnothing$, and $w:\mathcal{C}\to\mathbb{C}$ is defined by
\begin{equation}
    w_\gamma \coloneqq \prod_{u\in\bar{\gamma}}\frac{a_{u,\pi(\gamma(u))}}{a_{u,\pi(u)}}. \notag
\end{equation}
For each $u\in[n]$, let $\mathcal{C}_u\coloneqq\{\gamma\in\mathcal{C}\mid u\in\bar{\gamma}\}$ denote the set of cycles that contain $u$.

The following lemma expresses the permanent of $A$ in terms of the partition function of its cycle polymer model.

\begin{lemma}
    \label{lemma:PolymerModelRepresentation}
    Let $A=(a_{u,v})\in\mathbb{C}^{n \times n}$ be a matrix, let $\pi \in S_n$ be a permutation such that the entries $a_{u,\pi(u)}$ are non-zero for all $u\in[n]$, and let $\mathcal{P}=(\mathcal{C},w,\sim)$ be the cycle polymer model of $A$ with respect to $\pi$. Then
    \begin{equation}
        \operatorname{per}(A) = \left(\prod_{u\in[n]}a_{u,\pi(u)}\right)Z_\mathcal{P}. \notag
    \end{equation}
\end{lemma}

\begin{proof}
    By the definition of the permanent and its invariance under permutations,
    \begin{equation}
        \operatorname{per}(A) = \sum_{\sigma \in S_n}\prod_{u\in[n]}a_{u,\sigma(u)} = \sum_{\sigma \in S_n}\prod_{u\in[n]}a_{u,\pi(\sigma(u))}. \notag
    \end{equation}
    Since $a_{u,\pi(u)}$ is non-zero for all $u\in[n]$,
    \begin{equation}
        \operatorname{per}(A) = \left(\prod_{u\in[n]}a_{u,\pi(u)}\right)\sum_{\sigma \in S_n}\prod_{u\in[n]}\frac{a_{u,\pi(\sigma(u))}}{a_{u,\pi(u)}}. \notag
    \end{equation}
    For a permutation $\sigma \in S_n$, let $\Gamma_\sigma$ denote the set of cycles of $\sigma$. By grouping the factors according to the cycles in $\Gamma_\sigma$,
    \begin{equation}
        \operatorname{per}(A) = \left(\prod_{u\in[n]}a_{u,\pi(u)}\right)\sum_{\sigma \in S_n}\prod_{\gamma\in\Gamma_\sigma}\prod_{u\in\bar{\gamma}}\frac{a_{u,\pi(\gamma(u))}}{a_{u,\pi(u)}}. \notag
    \end{equation}
    Note that the cycles of length $1$ contribute a factor of $1$, while each cycle of length at least $2$ is a polymer $\gamma\in\mathcal{C}$ and contributes a factor of $w_\gamma$. By identifying each permutation $\sigma \in S_n$ with its set of cycles of length at least $2$, which is an admissible set of polymers, we obtain
    \begin{equation}
        \operatorname{per}(A) = \left(\prod_{u\in[n]}a_{u,\pi(u)}\right)\sum_{\Gamma\in\mathcal{G}}\prod_{\gamma\in\Gamma}w_\gamma = \left(\prod_{u\in[n]}a_{u,\pi(u)}\right)Z_\mathcal{P}. \notag
    \end{equation}
    This completes the proof.
\end{proof}

We now express the logarithm of the partition function of the cycle polymer model as a sum over closed walks. A \emph{closed walk} from $u\in[n]$ is a sequence of at least two elements of $[n]$ that starts and ends with $u$, and whose consecutive elements are distinct. Let $\mathcal{W}$ denote the set of all closed walks, and, for each $u\in[n]$, let $\mathcal{W}_u$ denote the set of closed walks from $u$.

The \emph{loop erasure} of a closed walk $W$ is the ordered tuple $\Gamma_W$ of cycles obtained as follows. Initialise a sequence $\rho$ containing the first element of $W$ and an empty tuple $\Gamma_W$. For each subsequent element $v$ of $W$ in order, if $v$ is an element of $\rho$, then let $\gamma$ denote the cycle with cycle notation given by the elements of $\rho$ starting from $v$, append $\gamma$ to $\Gamma_W$, and remove the elements of $\rho$ after $v$; otherwise, append $v$ to $\rho$. We denote by $\gamma_W$ the last cycle in $\Gamma_W$. Note that each cycle in $\Gamma_W$ is a polymer in $\mathcal{C}$, since consecutive elements of $W$ are distinct. Further, each step of $W$ belongs to exactly one cycle in $\Gamma_W$, and hence $\prod_{\gamma\in\Gamma_W}w_\gamma$ is the product of $a_{v,\pi(x)}/a_{v,\pi(v)}$ over all steps of $W$ from $v$ to $x$.

The following lemma, which is a special case of a result of Viennot~\cite{viennot1986heaps} (see Ref.~\cite[Theorem~A.5]{helmuth2016loop}), establishes that loop erasure is a bijection between closed walks and pyramids.

\begin{lemma}
    \label{lemma:LoopErasure}
    Let $u\in[n]$. Then the map sending each closed walk $W\in\mathcal{W}_u$ to the heap of $\Gamma_W$ is a bijection onto the set of pyramids in $\mathcal{G}_\mathrm{P}$ whose last polymer is in $\mathcal{C}_u$.
\end{lemma}

We assign to each closed walk its contribution to the pyramid expansion of the cycle polymer model. Define $\Phi:\mathcal{W}\to\mathbb{C}$ by
\begin{equation}
    \Phi(W) \coloneqq \frac{(-1)^{\abs{\Gamma_W}-1}}{\abs{\Gamma_W}\abs{\gamma_W}}\prod_{\gamma\in\Gamma_W}w_\gamma. \notag
\end{equation}

The following lemma expresses the pyramid expansion of the cycle polymer model of a permutationally-dominant matrix as a sum over $\mathcal{W}$.

\begin{lemma}
    \label{lemma:WalkRepresentation}
    Let $\lambda>1$, let $A=(a_{u,v})\in\mathbb{C}^{n \times n}$ be a $\lambda$-permutationally-dominant matrix with respect to a permutation $\pi \in S_n$, and let $\mathcal{P}=(\mathcal{C},w,\sim)$ be the cycle polymer model of $A$ with respect to $\pi$. Then the pyramid expansion converges absolutely to a logarithm of $Z_\mathcal{P}$, given by
    \begin{equation}
        \log(Z_\mathcal{P}) = \sum_{W\in\mathcal{W}}\Phi(W). \notag
    \end{equation}
    Further, $\sum_{W\in\mathcal{W}}\abs{\Phi(W)}\leq\frac{n}{2\lambda(\lambda-1)}$.
\end{lemma}

\begin{proof}
    By $\lambda$-permutational dominance, for all $v\in[n]$,
    \begin{equation}
        \sum_{x\in[n]{\setminus}\{v\}}\frac{\abs{a_{v,\pi(x)}}}{\abs{a_{v,\pi(v)}}} \leq \frac{1}{\lambda}. \notag
    \end{equation}
    Therefore,
    \begin{equation}
        \sum_{W\in\mathcal{W}}\abs{\Phi(W)} \leq \frac{1}{2}\sum_{u\in[n]}\sum_{W\in\mathcal{W}_u}\prod_{\gamma\in\Gamma_W}\abs{w_\gamma} \leq \frac{n}{2}\sum_{k=2}^\infty\lambda^{-k} = \frac{n}{2\lambda(\lambda-1)}, \notag
    \end{equation}
    where the second inequality follows since, for all $u\in[n]$ and all $k\geq2$, the closed walks in $\mathcal{W}_u$ of length $k$ contribute at most $\lambda^{-k}$. By Lemma~\ref{lemma:LoopErasure}, each pyramid in $\mathcal{G}_\mathrm{P}$ with last polymer $\gamma$ is the heap of $\Gamma_W$ for exactly $\abs{\gamma}$ closed walks $W\in\mathcal{W}$. Further, for all $W\in\mathcal{W}$, the heap of $\Gamma_W$ has $\abs{\Gamma_W}$ polymers and last polymer $\gamma_W$. Therefore, by grouping the closed walks in $\mathcal{W}$ according to their pyramids,
    \begin{equation}
        \sum_{H\in\mathcal{G}_\mathrm{P}}\frac{1}{\abs{H}}\prod_{\gamma\in H}\abs{w_\gamma} = \sum_{W\in\mathcal{W}}\abs{\Phi(W)} < \infty. \notag
    \end{equation}
    Hence, the pyramid expansion converges absolutely, and so, by the formal identity $\exp(\log(Z_\mathcal{P}))=Z_\mathcal{P}$, it converges to a logarithm of $Z_\mathcal{P}$. Further, by the same grouping,
    \begin{equation}
        \log(Z_\mathcal{P}) = \sum_{H\in\mathcal{G}_\mathrm{P}}\frac{(-1)^{\abs{H}-1}}{\abs{H}}\prod_{\gamma\in H}w_\gamma = \sum_{W\in\mathcal{W}}\Phi(W). \notag
    \end{equation}
    This completes the proof.
\end{proof}

\section{Approximation Algorithms}
\label{section:ApproximationAlgorithms}

In this section we establish a fully polynomial randomised approximation scheme for the permanent of permutationally-dominant matrices based on the polymer model representation of the permanent and importance sampling of the pyramid expansion.

\subsection{Walk Estimator}
\label{section:WalkEstimator}

In this section we introduce the walk estimator, an algorithm for estimating $\log(Z_\mathcal{P})$ by sampling closed walks.

\begin{definition}[Walk estimator]
    Let $\lambda>1$, and let $A=(a_{u,v})\in\mathbb{C}^{n \times n}$ be a $\lambda$-permutationally-dominant matrix with respect to a permutation $\pi \in S_n$. The \emph{walk estimator} is the following randomised algorithm:
    \begin{enumerate}
        \item Choose $u\in[n]$ uniformly at random. Initialise a sequence $\rho=(u)$ and an empty tuple $\Gamma$, and set $v=u$ and $X=0$.
        \item Choose one of the following:
        \begin{enumerate}
            \item For each $x\in[n]{\setminus}\{v\}$, with probability $\abs{a_{v,\pi(x)}}/\abs{a_{v,\pi(v)}}$, set $v=x$.
            \item With probability $1-\sum_{x\in[n]{\setminus}\{v\}}\abs{a_{v,\pi(x)}}/\abs{a_{v,\pi(v)}}$, return $X$.
        \end{enumerate}
        \item Choose one of the following:
        \begin{enumerate}
            \item If $v=u$, let $\gamma$ denote the cycle whose cycle notation is $\rho$. Append $\gamma$ to $\Gamma$, set $\rho=(u)$, and set
            \begin{equation}
                X = X+\frac{(-1)^{\abs{\Gamma}-1}n}{\abs{\Gamma}\abs{\gamma}}\prod_{\gamma'\in\Gamma}\frac{w_{\gamma'}}{\abs{w_{\gamma'}}}. \notag
            \end{equation}
            \item If $v$ is an element of $\rho$ other than $u$, let $\gamma$ denote the cycle with cycle notation given by the elements of $\rho$ starting from $v$. Append $\gamma$ to $\Gamma$, and remove the elements of $\rho$ after $v$.
            \item Otherwise, append $v$ to $\rho$.
        \end{enumerate}
        \item Return to step 2.
    \end{enumerate}
\end{definition}

The following lemma bounds the expected runtime of the walk estimator, and establishes that it is an unbiased estimator of $\log(Z_\mathcal{P})$ with bounded second moment.

\begin{lemma}
    \label{lemma:WalkEstimator}
    Let $\lambda>1$, let $A=(a_{u,v})\in\mathbb{C}^{n \times n}$ be a $\lambda$-permutationally-dominant matrix with respect to a permutation $\pi \in S_n$, and let $\mathcal{P}=(\mathcal{C},w,\sim)$ be the cycle polymer model of $A$ with respect to $\pi$. Then the walk estimator has expected runtime $O(\lambda/(\lambda-1))$ after $O(n^2)$ preprocessing time. Further, its output $X$ satisfies
    \begin{equation}
        \mathbb{E}[X]=\log(Z_\mathcal{P}) \quad \text{and} \quad \mathbb{E}[\abs{X}^2]\leq\frac{n^2(\lambda^2+1)}{4(\lambda^2-1)^2}. \notag
    \end{equation}
\end{lemma}

\begin{proof}
    We first bound the expected runtime of the walk estimator. By $\lambda$-permutational dominance, for all $v\in[n]$,
    \begin{equation}
        \sum_{x\in[n]{\setminus}\{v\}}\frac{\abs{a_{v,\pi(x)}}}{\abs{a_{v,\pi(v)}}} \leq \frac{1}{\lambda} < 1, \notag
    \end{equation}
    and hence the probabilities in each iteration of the walk estimator are valid, and the walk estimator has at least $m$ iterations with probability at most $\lambda^{-(m-1)}$. Therefore, the expected number of iterations is at most
    \begin{equation}
        \sum_{m=1}^\infty\lambda^{-(m-1)} = \frac{\lambda}{\lambda-1}. \notag
    \end{equation}
    Each iteration can be implemented in $O(1)$ amortised time after $O(n^2)$ preprocessing via the alias method~\cite{walker1977efficient, vose1991linear}, since each element appended to $\rho$ is removed at most once. In particular, we record the positions of the elements of $\rho$ in an array shared between runs, and maintain the product of the weights of the cycles in $\Gamma$. The total expected runtime of the walk estimator is therefore $O(\lambda/(\lambda-1))$.

    We next show that $X$ is an unbiased estimator of $\log(Z_\mathcal{P})$. Note that the values taken by $v$ form a random walk from $u$, and that $\rho$ and $\Gamma$ are updated according to its loop erasure. Hence, each term added to $X$ corresponds to a closed walk $W\in\mathcal{W}_u$ formed by the values taken by $v$ so far, and is given by $n\Phi(W)/\prod_{\gamma\in\Gamma_W}\abs{w_\gamma}$. Further, each closed walk $W\in\mathcal{W}$ with $\Phi(W)$ non-zero is formed with probability $\frac{1}{n}\prod_{\gamma\in\Gamma_W}\abs{w_\gamma}$, and so contributes $\Phi(W)$ to $\mathbb{E}[X]$. Since $\sum_{W\in\mathcal{W}}\abs{\Phi(W)}<\infty$, we may interchange the expectation and the sum over closed walks, and so, by Lemma~\ref{lemma:WalkRepresentation},
    \begin{equation}
        \mathbb{E}[X] = \sum_{W\in\mathcal{W}}\Phi(W) = \log(Z_\mathcal{P}). \notag
    \end{equation}

    Finally, we bound the second moment of $X$. Since each term added to $X$ requires at least two iterations, the walk estimator adds at least $m$ terms to $X$ with probability at most $\lambda^{-2m}$ for all $m\in\mathbb{Z}^+$. Therefore, since each term has absolute value at most $\frac{n}{2}$,
    \begin{equation}
        \mathbb{E}[\abs{X}^2] \leq \frac{n^2}{4}\sum_{m=1}^\infty(2m-1)\lambda^{-2m} = \frac{n^2(\lambda^2+1)}{4(\lambda^2-1)^2}. \notag
    \end{equation}
    This completes the proof.
\end{proof}

\subsection{Permanent Estimator}
\label{section:PermanentEstimator}

We now combine the walk estimator with the polymer model representation of the permanent to obtain our main result.

\begin{theorem}
    \label{theorem:PermanentAlgorithm}
    Fix $\lambda>1$. There is a fully polynomial randomised approximation scheme for the permanent of $\lambda$-permutationally-dominant matrices $A\in\mathbb{C}^{n \times n}$ with runtime $O(n^2\epsilon^{-2})$.
\end{theorem}

\begin{proof}
    Let $A=(a_{u,v})\in\mathbb{C}^{n \times n}$ be a $\lambda$-permutationally-dominant matrix with respect to a permutation $\pi \in S_n$. Since $\lambda>1$, for all $u\in[n]$, $\pi(u)$ is the unique $v\in[n]$ that maximises $\abs{a_{u,v}}$, and hence $\pi$ can be computed in $O(n^2)$ time. Let $\mathcal{P}=(\mathcal{C},w,\sim)$ be the cycle polymer model of $A$ with respect to $\pi$. By Lemma~\ref{lemma:WalkEstimator}, the walk estimator has expected runtime $O(1)$ after $O(n^2)$ preprocessing time, and its output $X$ satisfies
    \begin{equation}
        \mathbb{E}[X]=\log(Z_\mathcal{P}) \quad \text{and} \quad \mathbb{E}[\abs{X}^2]\leq\frac{n^2(\lambda^2+1)}{4(\lambda^2-1)^2}. \notag
    \end{equation}

    Let $\widehat{X}$ denote the mean of $N=\lceil6n^2\epsilon^{-2}(\lambda^2+1)/(\lambda^2-1)^2\rceil$ independent outputs of the walk estimator. By Chebyshev's inequality,
    \begin{equation}
        \Pr\left[\abs{\log(Z_\mathcal{P})-\widehat{X}}\geq\frac{\epsilon}{2}\right] \leq \frac{4\mathbb{E}[\abs{X}^2]}{N\epsilon^2} \leq \frac{1}{6}. \notag
    \end{equation}
    Let $\widehat{Z}\coloneqq\left(\prod_{u\in[n]}a_{u,\pi(u)}\right)e^{\widehat{X}}$ denote our estimate of $\operatorname{per}(A)$. By Lemma~\ref{lemma:PolymerModelRepresentation} and Lemma~\ref{lemma:WalkRepresentation}, if $\abs{\log(Z_\mathcal{P})-\widehat{X}}<\frac{\epsilon}{2}$, then
    \begin{equation}
        \abs{\operatorname{per}(A)-\widehat{Z}} = \abs{1-e^{\widehat{X}-\log(Z_\mathcal{P})}}\abs{\operatorname{per}(A)} \leq \left(e^{\epsilon/2}-1\right)\abs{\operatorname{per}(A)} \leq \epsilon\abs{\operatorname{per}(A)}, \notag
    \end{equation}
    where the first inequality follows from $\abs{1-e^z} \leq e^{\abs{z}}-1$ for all $z\in\mathbb{C}$. Hence, $\widehat{Z}$ is a multiplicative $\epsilon$-approximation to $\operatorname{per}(A)$ with probability at least $\frac{5}{6}$. The expected runtime of computing $\widehat{Z}$ is at most $\tau=O(n^2\epsilon^{-2})$. By Markov's inequality, the runtime exceeds $6\tau$ with probability at most $\frac{1}{6}$. Running the computation for time at most $6\tau$ therefore yields an algorithm with runtime $O(n^2\epsilon^{-2})$ whose output is a multiplicative $\epsilon$-approximation to $\operatorname{per}(A)$ with probability at least $\frac{2}{3}$. This completes the proof.
\end{proof}

\section{Conclusion \& Outlook}
\label{section:ConclusionOutlook}

We have established a fully polynomial randomised approximation scheme for the permanent of $\lambda$-permutationally-dominant matrices for any fixed $\lambda>1$, with runtime $O(n^2\epsilon^{-2})$. Our approach is based on a polymer model representation of the permanent in which the polymers are cycles, and importance sampling of the logarithm of its partition function, which we express as a sum over closed walks via the theory of heaps of pieces of Viennot.

There are several natural directions for future research. Our algorithm is randomised, and it remains open whether there is a deterministic polynomial-time algorithm for this class of matrices. Further, it would be interesting to determine whether the condition $\lambda>1$ is optimal, in the sense that approximating the permanent is \#P-hard for $\lambda\leq1$. Finally, our approach may extend to other abstract polymer models whose pyramids can be sampled efficiently in proportion to the absolute values of their weights, and it would be interesting to identify such models.

\section*{Acknowledgements}

RLM was supported by the ARC Training Centre for Future Leaders in Quantum Computing (FLiQC), project number IC240100009.

\bibliography{bibliography}

\end{document}